\documentclass[preprint2]{aastex6}
\usepackage{graphicx}   % needed for figures
\usepackage{dcolumn}    % needed for some tables
\usepackage{bm}         % for math
\usepackage{verbatim}   % for math
\usepackage{booktabs}
\usepackage{amsmath,amssymb}
\usepackage{color}
\usepackage{floatrow}
\usepackage{multirow}
\usepackage{inputenc}

\bibpunct{(}{)}{;}{a}{}{,}
\usepackage{floatrow}
\usepackage[caption=false]{subfig}
\usepackage{savesym}
\savesymbol{tablenum}
\usepackage{siunitx}
\restoresymbol{SIX}{tablenum}
\usepackage{enumitem}   % custom enumerate labels

\usepackage[normalem]{ulem}
\newcommand\bsout{\bgroup\markoverwith{\textcolor{blue}{\rule[0.5ex]{2pt}{0.4pt}}}\ULon}

\begin{document}
\title{Stochastic Turbulent Acceleration in a fractal environment} \footnote{Accepted for publication in The Astrophysical Journal Letters} 
\author{Nikos Sioulas,  Heinz Isliker, and Loukas Vlahos}
\affiliation{Department of Physics, \; Aristotle University of Thessaloniki\\
GR-52124 Thessaloniki, Greece}

\date{\today}
\begin{abstract}
We analyze the stochastic acceleration of particles inside a fully developed turbulent plasma. It is well known that large-amplitude magnetic fluctuations and coherent structures in such an environment obey a fractal scaling,
and our specific aim is to study for the first time the effects of the fractality of these environments on stochastic acceleration.
We have shown that an injected Maxwellian energy distribution is heated and forms a high energy tail in a very short time. Using standard parameters for the low solar corona, the injected Maxwellian distribution of electrons gets heated from the initial  $100\,$eV to $10\,$KeV, and the power-law index of the high energy tail is about $-2.3$. The high energy tail starts around  $100\,$ keV, and reaches $10\,$MeV. The index of the power-law tail depends on the system size, and it is in good agreement with observed values for realistic system sizes.
The heating and acceleration process is very fast ($\sim 2\,$s). The reason why the acceleration time is so short is that the particles are trapped within small scale parts of the fractal environment, and their scattering mean free path reduces drastically. The 
%simultaneous 
presence of small scale activity also pulls easily particles from the thermal pool, so there is no need for a seed population. The mean square displacement in space and energy is superdiffusive for the high energy particles. 
\end{abstract}
\maketitle
%-----------------------------------------------------------------------------------
%-----------------------------------------------------------------------------------
\section{Introduction}\label{intro}
Magnetic reconnection, weak turbulence, and shock waves surrounded by passive scattering centers upstream and downstream were for years the prominent acceleration mechanisms in most astrophysical and laboratory plasmas \citep{Melrose2009,Melrose94}. Recent magnetohydrodynamic (MHD) and kinetic simulations, as well as analytical work have shown that magnetic reconnection can lead to self-generated turbulence  \citep{Matthaeus86, Onofri06, Drake06, Daughton11, Oishi15,  Isliker19}, driven strong turbulence can also host reconnecting and non-reconnecting current sheets \citep{Biskamp89, Lazarian99,  Biskamp00, Arzner06,  Servidio11,  Isliker17a},  and in shock waves turbulent reconnection will be present mainly downstream  \citep{Matsumoto15, leRoux16, Garrel18}. Similarly, coherent structures including reconnecting current sheets are now established to be key components of turbulence in magnetized plasmas \citep{Matthaeus11, Cargill12,  Karimabadi13a, KarimabadiLazarian2013, Karimabadi2014, Vlahos19}.  In most explosive space, astrophysical, or laboratory plasmas, e.g flares,  unstable astrophysical flows (solar wind and astrophysical jets), or large scale shocks (bow shock, Heliospheric termination shock, coronal mass ejections, supernova remnants), the heating and acceleration of particles is due to the synergy of large-amplitude magnetic disturbances (stochastic energization) and magnetic reconnection and/or shocks (systematic energization) \citep{Pisokas18, Comisso18, Comisso19}. 

Acceleration of particles inside fully developed MHD turbulence is a very complex problem and depends on many important factors: (1) The nature of the interaction of particles with the ``scattering centers''  can be stochastic, systematic or synergy of both. The scattering centers inside fully developed turbulence are either large-amplitude magnetic disturbances or coherent structures (current sheets or shocks); (2) the scaling properties of the scattering centers control the energy and space transport and play a crucial role in the acceleration time and the escape time inside the finite acceleration volume.

The processes put forward by Fermi at the beginning of the '50s to describe particle acceleration inside fully developed turbulence are very broad in nature and include the well-known (i) stochastic (second-order Fermi) \citep{Fermi49} and/or (ii) the systematic (first-order Fermi) process \citep{Fermi54}. One can explore these processes by using the concept of a random walk inside a network of scattering centers \citep{Manolakou99, Vlahos04, Arzner04, Onofri06, Turkmani05,  Vlahos16, Pisokas16, Isliker2017, Pisokas18,  Garrel18, Sioulas20}.

\cite{Fermi49} used several simplified assumptions in his analysis of the stochastic interaction of cosmic rays with large amplitude MHD fluctuations: (1) The interaction of a particle with the large amplitude magnet fluctuations (``magnetic clouds'')   is stochastic and the energy gain ($\delta W$) is given by the relation  \citep{Longair11}
\begin{equation} \label{energygain}
    \frac{\delta W}{W} \sim \frac{2}{c^2}(V^2-\vec{V} \cdot \vec{u}) ,
\end{equation}
where $\vec V$ is the characteristic velocity of the magnetic disturbance, $\vec{u}$ is the velocity of the charged particle and $c$ the speed of light. If  $\vec{V}\cdot \vec{u} < 0$ the particles gain energy, and if $\vec{V}\cdot \vec{u} > 0$ the particles lose energy.
(2) The scattering centers are uniformly distributed in space, and the interaction of the particles with the scattering centers is expected to follow Gaussian statistics. The particles execute a random walk with a characteristic mean free path $\lambda_{sc} $ between the scattering centers. The acceleration time $t_{acc}$ was estimated by the relation 
$t_{acc} \sim [(3c)/ (4V^2)] \lambda_{sc}.$
(3) The Fokker-Plank transport equation was used as the basic tool for the study of the statistical evolution of the particles. The transport coefficients were grossly simplified. The acceleration time $t_{acc}$   is a measure of the energy transport of the particles inside the acceleration volume, and the escape time $t_{esc}$ a measure of the transport properties in space,
    $t_{esc} \sim L^2/ D$,
where $L$ is the characteristic length of the acceleration volume and $D\sim \lambda_{sc} c$ is the spatial diffusion coefficient if the particles are following a  random walk between the scattering centers. Here $t_{acc}/t_{esc} \sim [c^2 / V^2 L^2] \lambda_{sc}^2$, 
therefore the power-law index of the high energy tail in the distribution function is strongly dependent on the mean free path $\lambda_{sc}$. The acceleration time and the escape time are functions of  $\lambda_{sc}$ and are estimated using the assumptions listed above. 

Independently from Fermi's treatment and assumptions, it can be shown that the steady state solution of the energy continuity equation inside a finite acceleration volume (leaky box approximation) for stochastic Fermi acceleration  (Eq.\ \ref{energygain}) is 
\begin{equation}
f(W) \sim W^{-(1+t_{acc}/t_{esc})}    
\label{Fermislaw}
\end{equation}
\citep{Longair11}.

The results and the simplifications listed above for the stochastic Fermi acceleration have been questioned recently \citep{Pisokas16, Sioulas20}: Both the transport properties in space and energy are not normal and the interactions of the particles with the scatterers follow non-Gaussian statistics.

It is well known that 
%the scaling properties of 
large scale magnetic disturbances and coherent structures in fully developed MHD turbulence follow monofractal or multifractal %properties, 
scalings,
both in space and laboratory plasma \citep{Tu95, Marsch97, Shivamoggi97, Biskamp03, Dimitropoulou13, Leonardis13,  Schaffner15, Isliker19}. \cite{Dimitropoulou09} examined the relationship between the fractal properties of the photospheric magnetic patterns and those of the coronal magnetic field discontinuities (current sheets) in solar active regions. \cite{Isliker19} analyzed the current fragmentation of a large scale current sheet formed during magnetic flux emergence on the Sun and show that the fragments have a fractal structure, with a fractal dimension $D_F= 1.7-1.8.$

After all, in fully developed turbulence the coherent structures and the large amplitude magnetic fluctuations are located on a fractal set with dimension $D_F$, and the mean free path of the particles with the scattering centers $(\lambda_{sc})$ is not a simple constant \citep{Isliker03}.

In this article, we explore for the first time stochastic Fermi acceleration when the large amplitude MHD magnetic fluctuations have a fractal structure in space, and the particles are executing a random walk in this environment. In section 2, we briefly outline the essential characteristics of the random walk in a fractal environment. In section 3, we present our Monte Carlo simulation model, and in section 4, we analyze our results. In the final section, we discuss the implication of our results for turbulent stochastic Fermi acceleration.

\section{Random walk in a fractal environment}{\label{rand_walk}}

\begin{figure}[!ht]
     \begin{center}
         \includegraphics[width=0.7\textwidth]{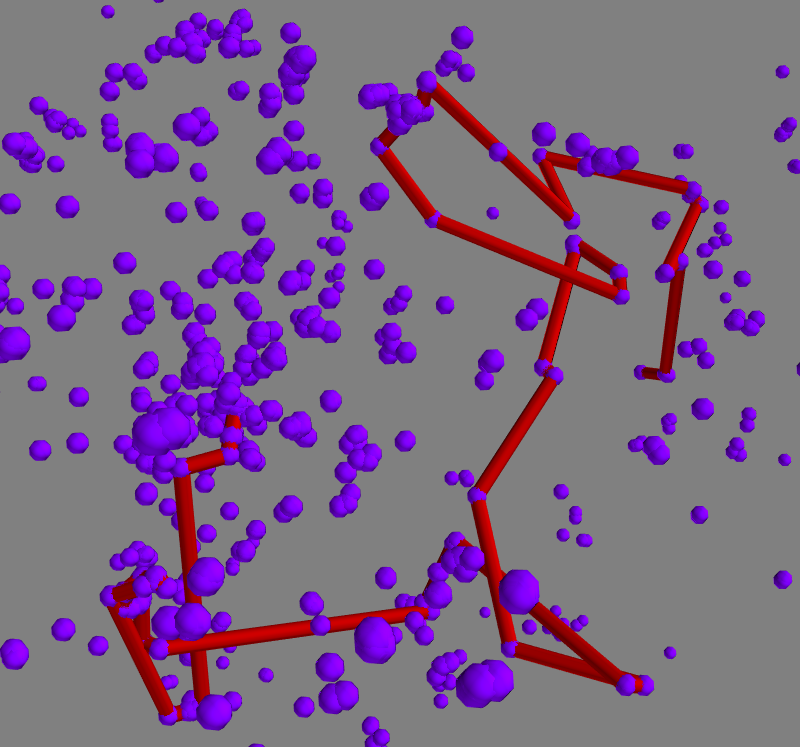}
     \caption {Illustration of the random walk through a fractal environment: Part of the fractal, with its constituent elementary volumes in blue color, and the orbit of a particle in red color, moving along straight paths and occasionally scattering off elementary volumes of the fractal. }\label{orbit2}
     \end{center}
     \end{figure}

\cite{Isliker03} have analyzed the random walk in the environment of a natural fractal, where the fractal is embedded in 3D space and the particles move freely in the empty space not occupied by the fractal until they occasionally collide with parts of the fractal set, where they undergo some kind of scattering. The particles thus move across the fractal, not along it. The fractal is natural in the sense that it is made up of small elementary and finite volumes (and not of points, line-segments, etc., as in the case of mathematical fractals), and it also is of finite, usually though large size, such that a clear fractal scaling holds from the fractal's size down to the size of its elementary volumes. The nature of the random walk is illustrated in Fig.\ \ref{orbit2}.

\cite{Isliker03} derived the probability density function (pdf) $p_F(dr)$  of the distances $dr$ a particle travels in between subsequent encounters with the fractal,   assuming that initially, a particle resides on a part of the fractal and then moves freely into a random direction until it hits another part of the fractal. For fractals with fractal dimension $D_F$ less than 2 (the case of interest here), this pdf turns out to be of power-law form in good approximation,
\begin{equation}
p_F(r) = A\, dr^{D_F-3}  
\label{fracstep}
\end{equation}
with $A$ a normalization constant, which is a function of the size of the natural fractal and the size of the elementary volumes it is constituted of. With $D_F< 2$, it follows that the power-law index of $p_F(dr)$ lies in the range $-3< D_F-3 < -1$, which means that $p(dr)$  has the same asymptotic (large $dr$) functional form as the stable Levy distributions. Particles thus occasionally perform large spatial jumps or ``Levy flights", and spatial transport must be expected to be anomalous \citep{Vlahos08Tut}. A peculiarity of the pdf $p_F(dr)$ is that it is defective, i.e. it is normalized to a value less than one, which implies a finite probability for direct escape in one step, without any secondary encounter with the fractal.

\section{Our Model}
\label{sec:headings}
%todo 
We construct a 3D box of linear size $L=10^{10}\,$cm. We initiate the simulation by uniformly placing $10^{6}$ particles in the interior of the acceleration volume. At time $t=0$, the energy distribution of the particles is a Maxwellian with temperature $T$. We then allow each particle to perform a free flight of length $dr_{i}^{(j)}$, before it meets a scatterer (i.e.\ it undergoes an energization event), where it gains or loses energy stochastically according to Eq.\ \ref{energygain}. The scatterers in our model are assumed to form a fractal set of dimensions $D_{F}$=1.8 (see Sec.\ \ref{intro} and \ref{rand_walk}). From Eq.\ \ref{fracstep}, the probability density  $P(dr){\sim}dr^{-{\gamma}}$, with ${\gamma}=1.2$, yields the length of the spatial step dr$_{i}^{(j)}$ each particle performs. We assume that spatial steps range from ${\lambda_{sc}}_{min}=10^{2}\,$cm to ${\lambda_{sc}}_{max}=10^{10}\,$cm.  The turbulent volume is a multi-scale environment. The range of the steps used in this study covers the entire range from the kinetic to the MHD scale, the lower limit is of the order of several ion gyroradii, and the upper limit basically equals the size of the acceleration box. Our results are not sensitive to the exact values used for the lower and upper step limit, as long as it holds that ${\lambda_{sc}}_{min}<< {\lambda_{sc}}_{max}$.
 As a result, there are "long flights", where particles are carried in one step over large distances, in some cases almost through the entire system, before they encounter a scatterer.

\begin{figure}[!ht]
     \begin{center}
        %\includegraphics[width=0.49\textwidth]{fig/Spatial_distrib,Lmin=100MLim=10000000000nP=100000.png}
        %\sidesubfloat[]{
        %\includegraphics[width=0.45\textwidth]{f1a.png}}
        
           \includegraphics[width=0.8\textwidth]{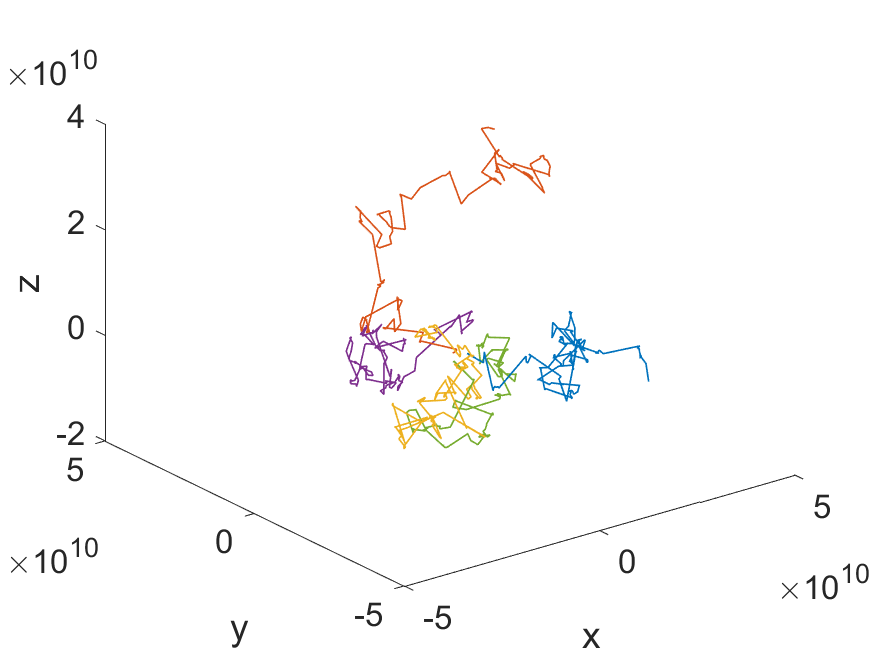}
           %\sidesubfloat[]

           \caption{Typical orbit for a number of particles, marked by different colors. Particles can be trapped inside regions of close-by scatterers or execute large flights.}\label{levyflights}
     \end{center}
     \end{figure}

 To completely specify the coordinates  of a particle each time it encounters a scatterer, we also generate a random number for the azimuthal angle ${\phi}$, $0 <{\phi}< 2{\pi} $, and one for $\cos({\theta})$, $-1 <\cos({\theta})< 1 $, with $\theta$ the polar angle. We then can determine the coordinates of each particle according to
 \begin{align*}
 x_{i}^{(j)}&=x_{i}^{(j-1)} + dr_{i}^{(j-1)}\cos{\phi} \sin{\theta}\\
 y_{i}^{(j)}
 &=y_{i}^{(j-1)} + dr_{i}^{(j-1)}\sin{\phi}\sin{\theta}\\
 z_{i}^{(j)}&= z_{i}^{(j-1)} + dr_{i}^{(j-1)}\cos{\theta}
 \end{align*} 
where $i=1,2,...,10^{6}$ is the particle index, and $j=1,2,...,N_{i}$ is the number of encounters a particle undergoes, %up to this point. 
%Finally, 
with $N_{i}$ being the total number of encounters each particle is subjected to before it  reaches the final simulation time or escapes from the acceleration volume.
 
During the free motion, the velocity of a particle remains constant, and,
since we know the length  $dr_{i}^{(j)}$ and the energy of the particles after an acceleration event, we can keep track of the time elapsed during the free flight as 
$dt_{i}^{(j)}=dr_{i}^{(j)}/|v_{i}^{(j)}|.$ Therefore, after a total number of $j$ encounters, the time elapsed for each particle is
${\tau}_{i}^{(j)}=\sum_{j=1}^{j}dt_{i}^{(j)}.$
We continue to keep track of the particles' energy and transport properties until they
reach the final simulation time or cross the boundaries of the box and, therefore, escape from the acceleration volume at time $t=t_{esc,i}$, which is, of course, different for each particle. In Fig.\ \ref{levyflights},  typical orbits in space are presented for a number of selected particles. Obviously, a standard orbit of the particles consists of a combination of long ``flights"
and efficient ``trapping" in localized spatial regions.

The conditions we simulate in this article are close to those found in the lower solar corona. We use as strength of the magnetic field $B = 100\,$G, as density of the plasma $n_{0} = 10^{9}\,$
cm$^{−3}$, and as ambient temperature  $T=100\,$ eV. The Alfven speed
is $V_{A}\sim   7 \times 10^{8}\,$cm/sec, a value close to the thermal speed of the electrons. With these parameters, the energy increments are close to ($\frac{{\delta}W}{W}$) $\sim (\frac{V_{A}}{c})^{2}\sim 10^{-4}$ (see Eq.\ \ref{energygain}).

%\section{Validation}

\begin{figure}[!ht]
     \begin{center}
     \sidesubfloat[]{
         \includegraphics[width=0.6\textwidth]{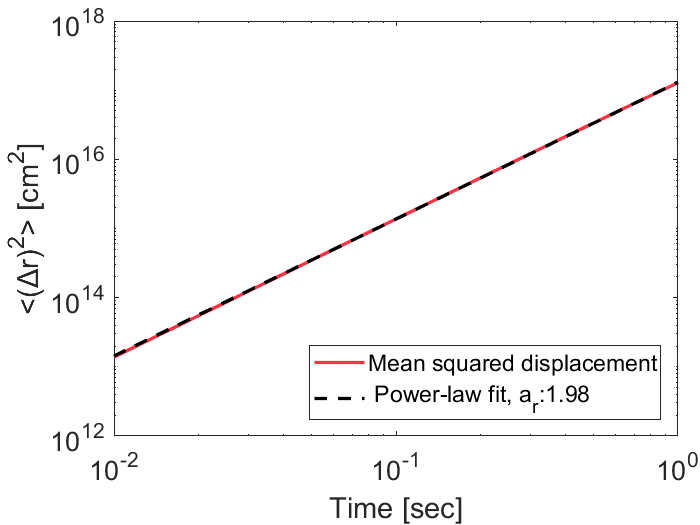}}\\
         \sidesubfloat[]{
         \includegraphics[width=0.6\textwidth]{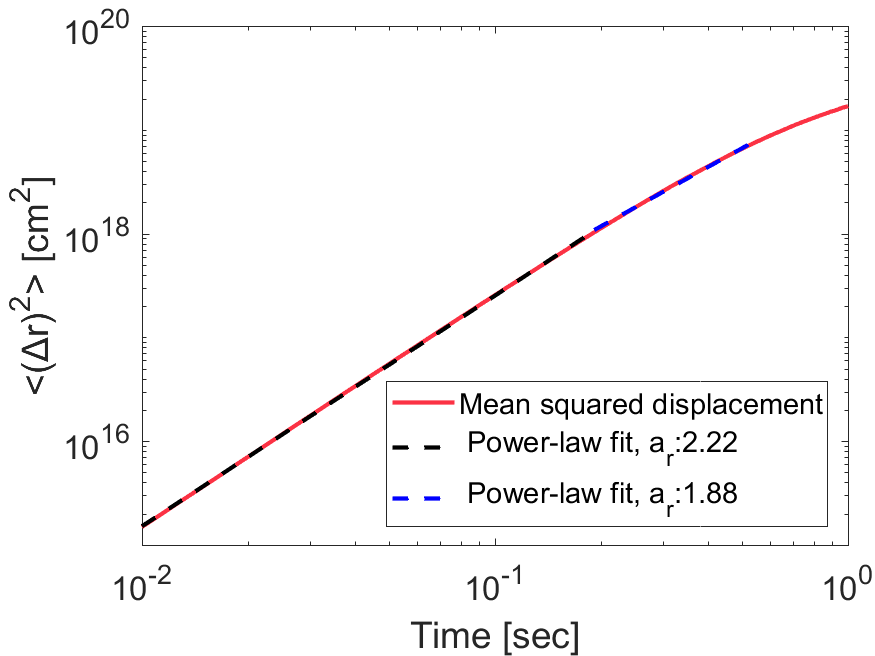}}
         \caption {(a) The mean square displacement of the particles as a function of time, in the absence of energization of the particles (passive scatterers).
        (b) The mean square displacement as a function of time, for the case where 
        the scatterers are active and energize the electrons.
}\label{rmsd}
     \end{center}
     \end{figure}

\section{Results}
\subsection{Spatial diffusion in the turbulent volume}\label{spatialdiff}
 In order to estimate the mean square displacement of the particles, we monitor their positions at prescribed and equi-spaced monitoring times $t^{n}$  ($n=1,...,N$). At time $t^{n}$ a particle's displacement from its initial position is ${{\Delta}\vec{r}_{i}^{n}}=\vec{r}^{n}_{i}-\vec{r}_{0i}$ and the mean square displacement for the ensemble of particles is
 \begin{equation} 
     {\langle}({\Delta}r^{n})^{2}{\rangle}=\frac{1}{N_{p}}\sum_{i=1}^{N_{p}}({\Delta}r^{n}_{i})^{2} . \label{msqd} 
 \end{equation}
 \par 
 We first assume that a particle's encounter with a scatterer solely influences its direction of motion, leaving its energy unchanged. In Fig.\ \ref{rmsd}a, we show the mean square displacement as a function of time. The diffusion for the particles interacting with the passive scatterers is ballistic, the scaling with time has a power-law index close to 2. This result agrees with the results obtained by \cite{Isliker03} (see Fig.\ 10 and Fig.\ 11 therein), where the particles also perform a random walk in an environment where a fractal with dimension $D_{F}<2$ resides.
 
We now turn to the case where the particles gain or lose energy stochastically through their interaction with active scatterers (see Eq.\ \ref{energygain}). 
The mean square displacement
%for which
%the spatial transport properties 
of the electrons 
%inside the acceleration  volume, 
is shown in Fig.\ \ref{rmsd}b,  it exhibits a superdiffusive scaling,  $<(\Delta r)^2> \sim t^{2.2}$,  
%We can also see that 
with the power-law index decreasing to $1.88$ after $t \sim 0.2\,$s.

     %In Fig. {\ref{rmsd}}b we illustrate the relation between the characteristic scaling index a$_r$ of the mean square displacement with the escape energy (i.e., the particles escape from the acceleration volume). To achieve this, we separate the particles in regards to their escape-energy, which is, of course, different for each particle, into (logarithmically equispaced) bins. We then estimate the power-law index for each one of the populations of particles using information from their travel history. To examine the reason that the power-law index of Fig. {\ref{rmsd_active}}a decreases after a few tenths of a second we separate the process into two parts. One from t = 0 up
%to t$\sim$ 0.2 sec and one for t= 0.2 sec to t=0.8 sec. After t=0.8-0.9 sec time, most of the particles have departed. In Fig. {\ref{rmsd_active}}a   We illustrate the process described above for times up to t=0.2 sec. We can see that for energies up to 10$^4$ eV, the transport of the electrons resembles the one achieved using passive scatterers. In contrast, for greater energies, the index a$_{r}$ gradually increases finally reaching a value close to a$_{r}$ ${\sim}$ 2.6. For the second phase of the acceleration process, and for times up to t=0.8 sec we can observe the power-law index for the thermal particles slightly decreasing to an average value close to a$_{r}$ ${\sim}$1.9.In contrast, the particles with escape energy above 10$^{4}$ eV exhibit a substantial decrease in regards to their index a$_{r}$, in comparison to the initial phase of the process.
\begin{figure}[!ht]
     \begin{center}
     \sidesubfloat[]{
         \includegraphics[width=0.61\textwidth]{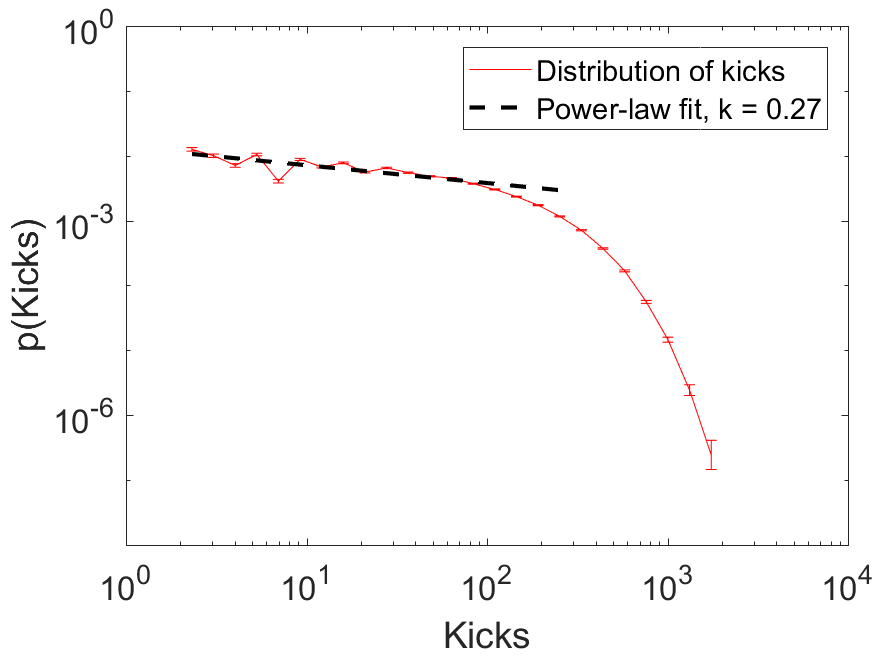}}\\
         \sidesubfloat[]{
          \includegraphics[width=0.6\textwidth]{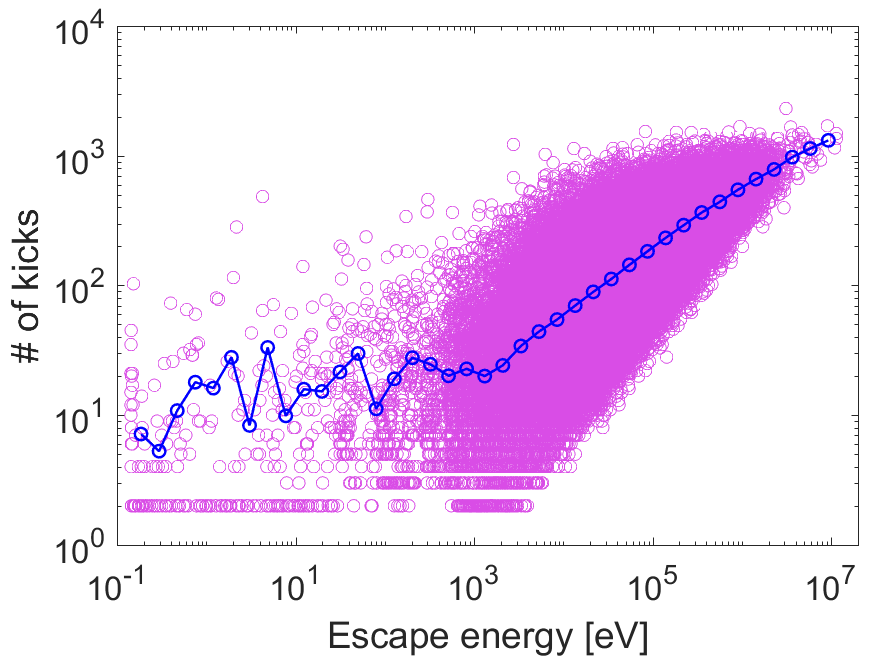} }\\
          %\sidesubfloat[]{
          %\includegraphics[width=0.7\textwidth]{f5a.png} }\\
          \sidesubfloat[]{
         \includegraphics[width=0.63\textwidth]{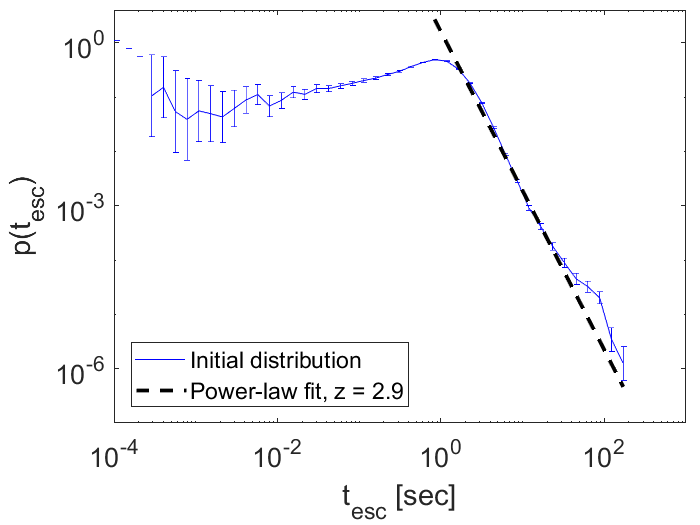} }
         \caption {(a) The distribution of the number of energization events (kicks) during the acceleration process in a fractal environment. (b) 
         Number of energization events 
         as a function of 
         the electron escape energy
         for each particle; the red line represents the binned median. 
         %of scatterings.  
         (c) The distribution of the electrons' escape times.}\label{kicks_hist}
     \end{center}
     \end{figure}
     
     \par 

In  Fig.\ {\ref{kicks_hist}}a we show the distribution of the total number of times the particles encounter a scatterer. 
%We can see that 
The number of encounters strongly varies, ranging from 2 to 2000, with a mean of $ \sim 140$ energization events per particle. From Fig.\ {\ref{kicks_hist}}b it is obvious that the particles trapped inside the acceleration volume are those accelerated most efficiently, yet only a fraction of the particles are subjected 
to a number of energization events that is high enough to be accelerated to super-thermal energies. 
%to a high enough number of energization events
%to be accelerated 
%that leads to acceleration 
%to super-thermal energies. 

%In {\ref{kicks_hist}}a a comparison is made between an environment where the scatterers are uniformly distributed across the volume (green line) and a fractal distribution of scatterers (blue line). Keeping all the other parameters constant, we can see that the particles traveling in an environment of uniformly distributed scatterers can experience a greater number of scatterings(and thus get accelerated to slightly higher energies) before leaving the acceleration volume. As discussed in Sec.\ref{rand_walk} the particles traveling in a fractal environment are allowed in some cases to perform very big spatial steps before encountering a scatterer. As a result, they can escape a lot faster from the acceleration volume having experienced on average a lower number of scatterings

%\begin{figure}[!ht]
    %\begin{center}
     %\includegraphics[width=0.49\textwidth]{f5a.png} 
         %\includegraphics[width=0.49\textwidth]{f5b.png} 
         %\includegraphics[width=0.45\textwidth]{t_esc,Lmin=1000000nP=10000.png} 
     %\caption {(a) Electron escape time as a function of the number of kicks, the red line represents the binned median of scatterings. (b)The distribution of the electrons escape time.}\label{tesc}
     %\end{center}
     %\end{figure}
%\par

The time spent by the electrons inside the acceleration volume is very important for our study, profoundly affecting, the power-law index of the kinetic energy distribution (see Sec.\ {\ref{intro}}). As Fig.\  \ref{kicks_hist}c shows,  most of the accelerated electrons escape from the volume quite early, while, for larger escape times, their distribution forms a power-law with index close to 2.9. The mean value of the escape times yields 
%is close to  
$t_{esc} \sim 1.9 \,$s. Comparing this result to \cite{Pisokas16}, where the acceleration process is taking place in an environment where the scatterers are uniformly distributed inside the acceleration volume, we observe a significant decrease in the escape time of the particles. 

 \subsection{Diffusion of electrons in energy space}

\begin{figure}[!ht]
     \begin{center}
     \sidesubfloat[]{
         \includegraphics[width=0.6\textwidth]{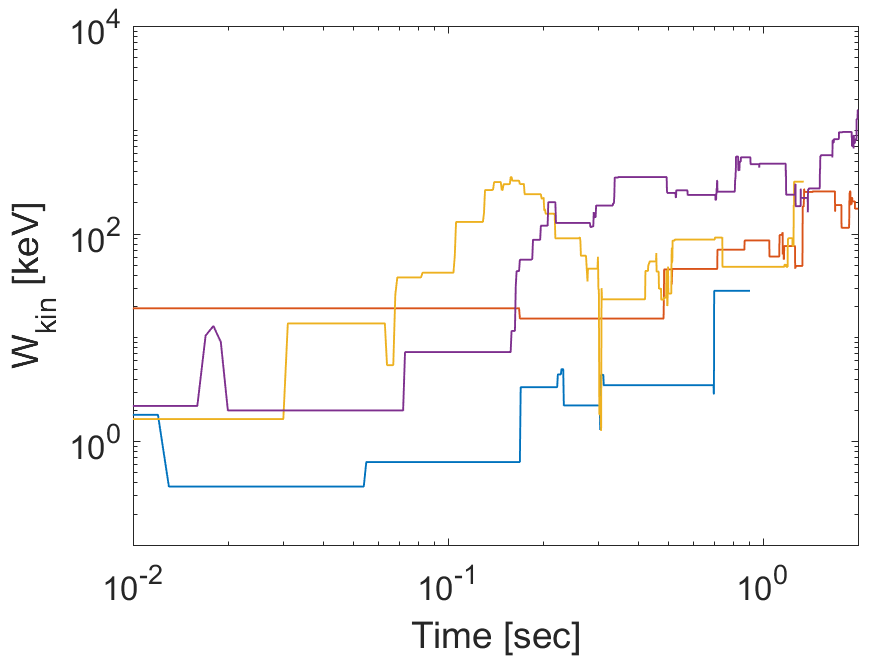}}\\
         \sidesubfloat[]{
         \includegraphics[width=0.6\textwidth]{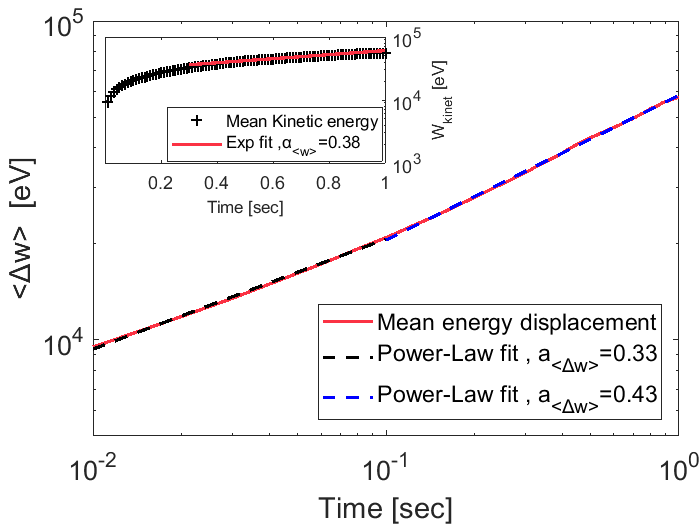}} \\
        \sidesubfloat[]{
          \includegraphics[width=0.6\textwidth]{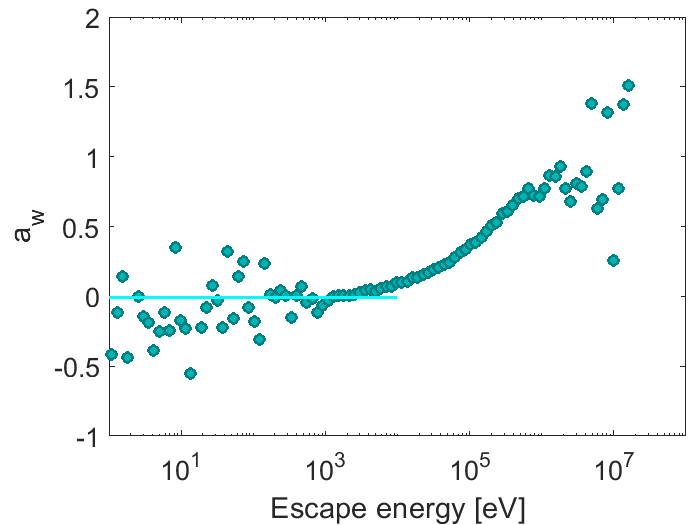}} %\includegraphics[width=0.49\textwidth]{W_esc_split/f6c_big6.png} 

     \caption {(a) Energization as a function of time for some typical electrons. (b) Mean displacement in the energy of the electrons as a function of time. The insert figure shows the mean kinetic energy of the electrons remaining inside the box as a function of time, together with an exponential fit (red). (c)
     Power-law index of the mean displacement in energy as a function of the escape energy.}\label{Wkinet}
     \end{center}
     \end{figure}
     
Equally important for our study are the transport properties of the kinetic energy of the energized particles. In an encounter with a scattering center, a particle (with index $i$) departs from the scatterer with renewed energy, 
$$W_i^{j+1}=W_i^j+\delta  W_i^j ,$$
where $\delta W_i^j$ is given by Eq.\ \ref{energygain}, and $j$ counts the number of energization events for the particle. 
%up to that point. 
In Fig.\ {\ref{Wkinet}}a, the 
%aforementioned 
energization process is presented for several typical particles, revealing its stochastic nature, but also exhibiting a slight predilection for encounters leading to energy gain.

Using the set of predefined monitoring times $t^n$ outlined in Sec.\ \ref{spatialdiff}, we keep track of the particles' energies $W^n_i$  at these times. If we denote by  $W_i^0$ the particles' initial energy, we can define the energy displacement as $\Delta  W_i^n=(W_i^n - W_i^0)$, and calculate the mean  displacement in energy  through the relation 
    \begin{equation}\label{enertr}
    <\Delta  W>(t^n)\equiv <\Delta  W^n>=\frac{1}{N_p}\sum_{i=1}^{i=N_p} \Delta W_i^n  ,
    \end{equation}
while the mean square displacement in energy is given by
    \begin{equation}\label{sqenergytr}
    <(\Delta  W)^2>(t^n) \equiv <(\Delta  W^n)^2>=\frac{1}{N_p}\sum_{i=1}^{i=N_p} (\Delta W_i^n)^2.
    \end{equation}
 
In general, we can assume that the mean energy displacement has a power-law form, $ <\Delta W>(t) = F_{W} t^{a_W}$, and the index $a_W$ can be estimated through a power-law fit.  Fig.\ \ref{Wkinet}b shows $<\Delta W>(t)$, there is indeed a power-law scaling with a slope $a_W \sim 0.33$ for times up to $0.1\,$s, and $a_W \sim 0.43$ for larger times.  The insert figure shows the evolution of the kinetic energy for the electrons remaining inside the acceleration box as a function of time. From the exponential fit we can estimate the acceleration time as $t_{acc} \sim 1/0.38 = 2.6\,$s \citep{Longair11}. Fig.\ \ref{Wkinet}c presents $a_W$ as a function of $W_{esc}$, from which it follows that there is no systematic acceleration for electrons with escape energy smaller than $10^4\,$eV. 
For the high energy particles, the scaling index gradually increases with energy, reaching a value close to $a_W\sim 1.5$. 

\begin{figure}[!ht]
     \begin{center}
     \sidesubfloat[]{
         \includegraphics[width=0.7\textwidth]{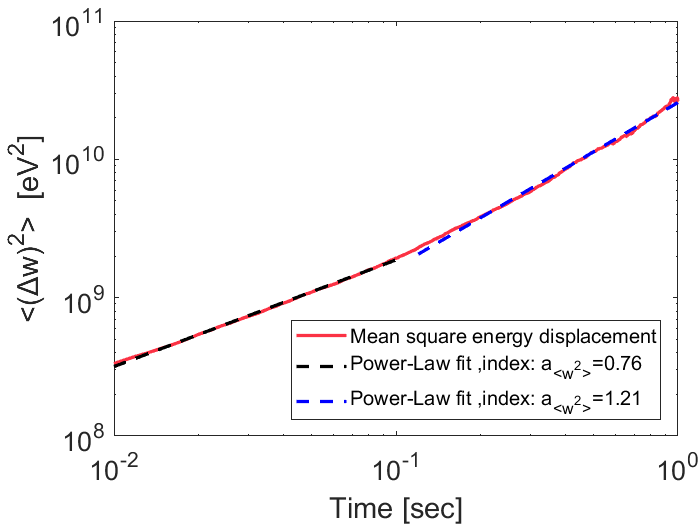}} \\
         \sidesubfloat[]{
         \includegraphics[width=0.675\textwidth]{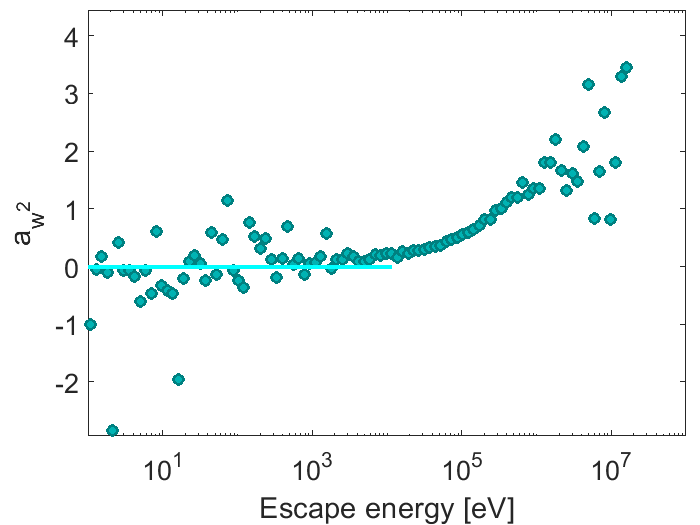}} 

     \caption {(a) Mean square displacement in energy as a function of of time. (b) Power-law index of the mean square displacement in energy as a function of the escape energy. }\label{Wkinet2}
     \end{center}
     \end{figure}

Similarly, in the case of the mean square displacement in energy we expect a power-law form $ <(\Delta W)^2>(t) = D_{W^2} t^{a_{W^2}}$.  In Fig.\ {\ref{Wkinet2}} the mean square  displacement in energy is presented. For times up to $t=0.1\,$s, the scaling is slightly sub-diffusive, following a power-law with index $a_{w^{2}}=0.76$. For larger times, the power-law index is $a_{w^{2}}=1.21$, indicating a super-diffusive behavior.
In Fig.\ \ref{Wkinet2}b we show $a_{W^2}$ as a function of the energy with which the electrons escape from the acceleration volume. As in the case of convective transport, electrons with energies smaller than $10\,$keV have on average a scaling index $a_w^{2}$ close to zero. For the super-thermal particles, we observe a substantial increase of the scaling index with increasing escape energy, moving from sub-diffusive to super-diffusive, even attaining values close to $a_{W^2}$=4 for the highest energy particles.
  
\begin{figure}[!ht]
     \begin{center}
     \sidesubfloat[]{
         \includegraphics[width=0.79\textwidth]{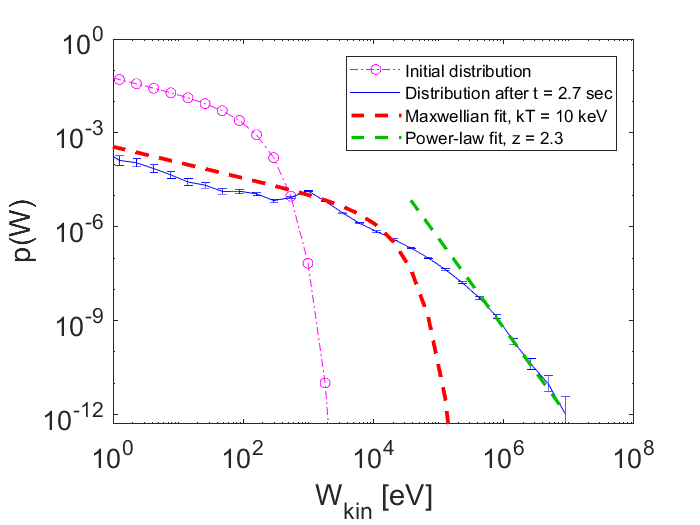}}\\
         \sidesubfloat[]{
       \includegraphics[width=0.7\textwidth]{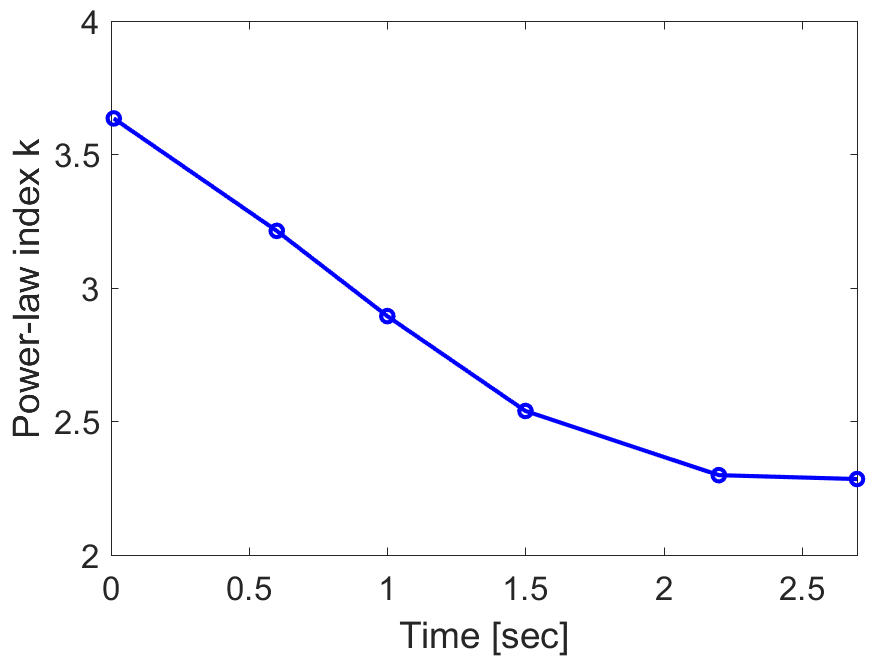}}\\ 
       %\sidesubfloat[]
       %{\includegraphics[width=0.75\textwidth]{f7c.png}}
     \caption { (a) Kinetic energy distribution at $t = 0$ and $t = 2.7\,$s (steady state) for the electrons remaining inside the box with size $L=10^{10}\,$cm, together with  a Maxwellian fit at low energies and a power-law fit at high energies.  (b) Temporal evolution of the power-law index of the kinetic energy distribution's tail.  
     %{\color{red} (c) Kinetic energy distribution at $t = 0$ and $t = 0.2\,$s  for the electrons remaining inside the box {\color{red} with size $L=10^{9}$ cm}, together with  a Maxwellian fit at low energies and a power-law fit at high energies.} 
     }\label{Wkinet3}
     \end{center}
     \end{figure}

%\begin{figure}[!ht]
     %\begin{center}
         %\includegraphics[width=0.49\textwidth]{f9a.png} 
      %\includegraphics[width=0.49\textwidth]{f9b.png} 
     %\caption { (a) The kinetic energy distribution
%of the electrons that have escaped from the box. (b) Electron escape energy %as a function escape time, the red line represents the binned median of scaterings}\label{Wkinet4}
     %\end{center}
    %\end{figure}   

In Fig.\ {\ref{Wkinet3}}a, we show the histogram of the kinetic energies for the particles that remain in the simulation box, normalized to unity, for the injected distribution and the 
%stabilized 
one at time $t = 2.7\,$s, along with a Maxwellian fit at low energies that yields a temperature  $T= 10\,$keV. In the first few milliseconds of the simulation, the low energy particles are actually already heated, and the high energy particles are already accelerated and form a power-law tail with index  $k \sim 3.8$. Fig.\ {\ref{Wkinet3}}b presents the evolution of the power-law index of the tail. After $2.7\,$s (which is equivalent to the acceleration time $t_{acc}$), the initially appearing power-law index $k \sim 3.8$ has decreased to an asymptotic value of about $k\sim 2.3$, the case shown in Fig.\ {\ref{Wkinet3}}a.

The power-law index of the tail of the kinetic energy distribution  can also be estimated through Fermi's expression  $k = 1 + t_{acc} /t_{esc} \sim 2.37$ (see Eq.\ \ref{Fermislaw}), which is  close  to the direct result from the power-law fit in Fig.\ {\ref{Wkinet3}}a. 
 
When reducing the size of the acceleration box, the particles do not have time to reach a steady-state distribution before escaping from the box, the slope of the  distribution at high energies becomes steeper and the maximum energy reached smaller. For example, for 
$L=10^9\,$cm the energy distribution remains the same in shape as the one for $L=10^{10}\,$cm in Fig.\ {\ref{Wkinet3}}, yet at an earlier time than in Fig.\ {\ref{Wkinet3}}a. Thus, the acceleration time becomes much shorter, the slope of the high energy tail gets steeper, $k\sim 3.4$ at $t=0.2\,$s (in complete accordance with Fig.\ {\ref{Wkinet3}}b), and the maximum energy reached is $1\,$MeV. 
%(see Fig.\  \ref{Wkinet3}c). 
Also, the heated Maxwellian distribution at low energies remains unaffected by a reasonable reduction of the acceleration volume (considering again earlier times than in Fig.\ {\ref{Wkinet3}}a). These results agree very well with the current observations from solar flares and space plasmas \cite{Oka2018}. When increasing the size of the simulation box above $10^{10}\,$cm, the energy distribution remains unaffected when comparing at equal times, since the energized particles are able to reach a steady-state.

According to \cite{Oka2018},  the observed index of the slope of the energetic particles is between $3$ and $5$ for most solar flares, which, based on our results, suggests that the acceleration box size is about $10^8-10^9$ cm.

\section{Summary and conclusions}

Stochastic turbulent acceleration and transport in space and astrophysical plasma has been analyzed so far with the use of the Fokker-Planck equation and the quasilinear approximation. Both approaches are appropriate for weak turbulence when the wave-particle interaction is a correct representation of the scattering of particles by the normal modes of an unstable plasma. Obviously, in strong and fully developed turbulence these approximations break down since the dominant acceleration mechanisms are large-amplitude magnetic disturbances and coherent structures (current sheets and shocks).  Following the initial suggestion by \cite{Fermi49}, we have explored the idea of particle acceleration and heating in the form of a random walk inside a network of scatterers. Fermi assumed that the scatterers (magnetic clouds) are uniformly distributed in space and the mean free path $\lambda_{sc}$ is constant. The mean free path plays a key role in the estimates of the acceleration and escape time and controls the power-law index of the high energy tail. As we outlined in the introduction, numerous numerical studies suggest that the spatial scaling of large-amplitude magnetic disturbances and coherent structures inside fully developed turbulence are located on a well defined fractal topology. We have explored here the role of the fractal scaling in stochastic Fermi acceleration.

The main results in this study are

\begin{enumerate}
    \item The stochastic interaction of particles with fractal large-amplitude magnetic fluctuations results in the heating and acceleration of particles. 
    \item The high energy particles are accelerated by a combination of intense trapping within small scale structures and delayed escape from the acceleration volume, undergoing up to thousands of energization events.
    
    \item The combined effects of trapping particles on small scales and of long ``flights''  dramatically affect the acceleration and escape time of stochastic acceleration. In particular, the acceleration time is strongly reduced when compared to acceleration in non-fractal, uniform environments.
    \item The spatial and energy transport of the high energy particles is superdiffusive. The Fokker-Planck equation for the study of the spatial and energy transport of high energy particles is inappropriate, it though is valid for thermal particles.
    \item The small scale interactions enhance the acceleration of particles from the thermal pool.
    \item We simulate in our study explosive phenomena (flares) in the low solar corona, using a simulation box with characteristic length $L=10^{10}$ cm. We have injected a very large number of electrons with a Maxwellian energy distribution with a  temperature of $100\,$eV.  In about two seconds, the energy distribution reaches an asymptotic shape, with a super-hot plasma with temperature $10\,$keV, and a power-law tail above $100\,$keV with power-law index $-2.3$, and reaching $10\,$MeV.
    \item When reducing the size of the box e.g.\ to $10^8-10^9\,$cm, the particles do not have time to reach a steady-state distribution before escaping from the system, and the power-law slope of the high energy tail becomes steeper, in agreement with the current observations from solar flares and space plasmas \citep{Oka2018}. 
    %The heated particles are not affected by a reasonable reduction of the acceleration volume. 
    Increasing the size of the acceleration box to $L>10^{10}\,$cm does not affect the energy distribution, since the particles in any case can reach a steady-state distribution.
   
\end{enumerate}{}

We confined our study to the stochastic Fermi acceleration of particles in a fractal turbulent environment, which turned out to be a very efficient and important mechanism for many turbulent astrophysical sources, beyond the case of solar flares studied here. Our next step is to incorporate %in the above study the 
coherent structures (reconnecting current sheets), as they are present in fully developed plasma turbulence.

\begin{acknowledgements}
We thank Theophilos Pisokas for his help in the initial phase of this project.
     \end{acknowledgements}
%-----------------------------------------------------------------------------------
%-----------------------------------------------------------------------------------
%\bibliographystyle{aasjournal}
%\bibliography{vlahosturb2}
\end{document}